\documentclass{llncs}

\usepackage{graphicx}
\usepackage{fullpage}

\newcommand{\G}{\mathcal{G}}
\newcommand{\X}{\mathcal{X}}
\newcommand{\Tree}{\mathcal{T}}
\newcommand{\F}{\mathcal{F}}
\newcommand{\levelanc}{{\mathit{level}\textrm{-}\mathit{ancestor}}}
\newcommand{\Rel}{\mathcal{R}}
\newcommand{\Lab}{\mathcal{L}}

\title{Improved Grammar-Based Compressed Indexes
\thanks{This work was partially supported by Google U.S./Canada PhD Fellowship and David
R. Cheriton Scholarships program (first author), and by Millennium
Institute for Cell Dynamics and Biotechnology (ICDB), Grant ICM
P05-001-F, Mideplan, Chile (second author).}}

\author{Francisco Claude \inst{1} \and Gonzalo Navarro \inst{2}}

\institute{ David R. Cheriton School of Computer Science, University
of Waterloo.  \email{fclaude@cs.uwaterloo.ca}.  \and Department of
Computer Science, University of Chile.
\email{gnavarro@dcc.uchile.cl}.}

\begin{document}

\maketitle

\begin{abstract}
We introduce the first grammar-compressed representation of a sequence
that supports searches in time that depends only logarithmically on
the size of the grammar. Given a text $T[1..u]$ that is represented by
a (context-free) grammar of $n$ (terminal and nonterminal) symbols and
size $N$ (measured as the sum of the lengths of the right hands of the
rules), a basic grammar-based representation of $T$ takes $N\lg n$
bits of space. Our representation requires $2N\lg n + N\lg u + \epsilon\,
n\lg n + o(N\lg n)$ bits of space, for any $0<\epsilon \le 1$.  It can
find the positions of the $occ$ occurrences of a pattern of length $m$
in $T$ in $O\left((m^2/\epsilon)\lg \left(\frac{\lg u}{\lg n}\right) +occ\lg n\right)$ time,
and extract any substring of length $\ell$ of $T$ in time
$O(\ell+h\lg(N/h))$, where $h$ is the height of the grammar tree.
\end{abstract}

\section{Introduction and Related Work}

Grammar-based compression is an active area of research that dates
from at least the seventies. A given sequence $T[1..u]$ over alphabet
$[1..\sigma]$ is replaced by a hopefully small (context-free) grammar
$\G$ that generates just the string $T$. Let $n$ be the number of
grammar symbols, counting terminals and nonterminals. Let $N$ be the
{\em size} of the grammar, measured as the sum of the lengths of the
right-hand sides of the rules. Then the grammar-compressed
representation of $T$ requires $N\lg n$ bits, versus the
$u\lg\sigma$ bits required by a plain representation.

Grammar-based methods can achieve universal compression \cite{KY00}.
Unlike statistical methods, that exploit frequencies to achieve
compression, grammar-based methods exploit repetitions in the text,
and thus they are especially suitable for compressing highly
repetitive sequence collections.  These collections, containing long
identical substrings, possibly far away from each other, arise when
managing software repositories, versioned documents, temporal
databases, transaction logs, periodic publications, and computational
biology sequence databases.

Finding the smallest grammar $\G^*$ that represents a given text $T$
is NP-complete \cite{Ryt03,CLLPPSS05}. Moreover, the smallest grammar
is never smaller than an LZ77 parse \cite{ZL77} of $T$. A simple
method to achieve an $O(\lg u)$-approximation to the smallest grammar
size is to parse $T$ using LZ77 and then to convert it into a grammar
\cite{Ryt03}. A more sophisticated approximation achieves ratio
$O(\lg(u/N^*))$, where $N^*$ is the size of $\G^*$.

While grammar-compression methods are strictly inferior to LZ77
compression, and some popular grammar-based compressors such as LZ78
\cite{ZL78}, Re-Pair \cite{LM00} and Sequitur \cite{NMWM04}, can
generate sizes much larger than the smallest grammar \cite{CLLPPSS05},
some of those methods (in particular Re-Pair) perform very well in
practice, both in classical and repetitive settings.\footnote{See the
  statistics in {\tt http://pizzachili.dcc.uchile.cl/repcorpus.html}
  for a recent experiment.}

In reward, unlike LZ77, grammar compression allows one to decompress
arbitrary substrings of $T$ almost optimally
\cite{GKPS05,BLRSSW11}. The most recent result \cite{BLRSSW11}
extracts any $T[p,p+\ell-1]$ in time $O(\ell+\lg u)$.  Unfortunately,
the representation that achieves this time complexity requires
$O(N\lg u)$ bits, possibly proportional but in practice many times
the size of the output of a grammar-based compressor. On the practical
side, applications like Comrad \cite{KBSCZ09} achieve good space and
time performance for extracting substrings of $T$.

More ambitious than just extracting arbitrary substring from $T$ is to
ask for indexed searches, that is, finding all the $occ$ occurrences
in $T$ of a given pattern $P[1..m]$. Self-indexes are compressed text
representations that support both operations, {\em extract} and {\em
  search}, in time depending only polylogarithmically on $u$.
They have appeared in the last decade \cite{NM07}, and have
focused mostly on statistical compression. As a result, they work well
on classical texts, but not on repetitive collections
\cite{MNSV09}. Some of those self-indexes have been adapted to
repetitive collections \cite{MNSV09}, but they cannot reach the
compression ratio of the best grammar-based methods.
 
Searching for patterns on grammar-compressed text has been faced
mostly in sequential form \cite{AB92}, that is, scanning the whole
grammar. The best result \cite{KMSTSA03} achieves time
$O(N+m^2+occ)$. This may be $o(u)$, but still linear in the size of
the compressed text. There exist a few self-indexes based on LZ78-like
compression \cite{FM05,ANScpm06.2,RO08}, but LZ78 is among the weakest
grammar-based compressors. In particular, LZ78 has been shown not to
be competitive on highly repetitive collections \cite{MNSV09}.

The only self-index supporting general grammar compressors \cite{CN09}
operates on ``straight-line programs'' (SLPs), where the right hands of the
rules are of length 1 or 2. Given such a grammar they achieve, among
other tradeoffs, $3n\lg n + n\lg u$ bits of space and
$O(m(m+h)\lg^2 n)$ search time, where $h$ is the height of the parse tree of
the grammar. A general grammar of $n$ symbols and
size $N$ can be converted into a SLP of $N-n$ rules.

More recently, a self-index based on LZ77 compression has been
developed \cite{KN11}. Given a parsing of $T$ into $n$ phrases, the
self-index uses $n\lg n + 2n\lg u + O(n\lg\sigma)$ bits of space,
and searches in time $O(m^2 h + (m+occ)\lg n)$, where $h$ is the nesting
of the parsing. Extraction requires
$O(\ell h)$ time.  Experiments on repetitive collections
\cite{CFMPN10,CFMPN11} show that the grammar-based compressor
\cite{CN09} can be competitive with the best classical self-index
adapted to repetitive collections \cite{MNSV09} but, at least that
particular implementation, is not competitive with the LZ77-based
self-index \cite{KN11}.

Note that the search time in both self-indexes depends on $h$.
This is undesirable as $h$ is only bounded by $n$.  That kind of 
dependence has been removed for extracting text substrings \cite{BLRSSW11}, 
but not for searches.

Our main contribution is a new representation of general context-free
grammars. The following theorem summarizes its properties. Note
that the search time is independent of $h$.

\begin{theorem} \label{thm:main}
Let a sequence $T[1..u]$ be represented by a context free grammar with
$n$ symbols, size $N$ and height $h$. Then, for any $0<\epsilon\le 1$,
there exists a data structure using at most $2N\lg n + N\lg u +
\epsilon\, n\lg n + o(N\lg n)$ bits of space that finds the $occ$
occurrences of any pattern $P[1..m]$ in $T$ in time
\linebreak $O\left((m^2/\epsilon)\lg \left(\frac{\lg u}{\lg n}\right) + occ\lg n\right)$. It can extract
any substring of length $\ell$ from $T$ in time
$O(\ell+h\lg(N/h))$. The structure can be built in $O(u+N\lg N)$ time
and $O(u\lg u)$ bits of working space.
\end{theorem}

In the rest of the paper we describe how this structure operates.
First, we preprocess the grammar to enforce several invariants useful
to ensure our time complexities. Then we use a data structure for
labeled binary relations \cite{CN09} to find the ``primary'' occurrences of
$P$, that is, those formed when concatenating symbols in the right
hand of a rule. To get rid of the factor $h$ in this part of the
search, we introduce a new technique to extract the first $m$ symbols
of the expansion of any nonterminal in time $O(m)$. To find the
``secondary'' occurrences (i.e., those that are found as the result of
the nonterminal containing primary occurrences being mentioned
elsewhere), we use a pruned representation of the parse tree of
$T$. This tree is traversed upwards for each secondary occurrence to
report. The grammar invariants introduced ensure that those traversals
amortize to a constant number of steps per occurrence reported. In
this way we get rid of the factor $h$ on the secondary occurrences
too.

\section{Basic Concepts}

\subsection{Sequence Representations}
\label{sec:seqs}

Our data structures use succinct representations of sequences. Given a 
sequence $S$ of length $N$, drawn from an alphabet of size $n$, we need
to support the following operations:

\begin{itemize}
\item $access(S,i)$: retrieves the symbol $S[i]$.
\item $rank_a(S,i)$: number of occurrences of $a$ in $S[1..i]$.
\item $select_a(S,j)$: position where the $j$th $a$ appears in $S$.
\end{itemize}

In the case where $n = 2$, Raman et al.~\cite{RRR02} proposed two
compressed representations of $S$, that are useful when the number $n'$
of 1s in $S$ is small (or large, which is not the case in this paper). 
One is called a ``fully indexable dictionary'' (FID). It takes
$n'\lg\frac{N}{n'}+O(n'+N\lg\lg N/\lg N)$ bits of space and supports all the 
operations in constant time. A weaker one is an ``indexable dictionary'' (ID), 
that takes $n'\lg\frac{N}{n'}+O(n'+\lg\lg N)$ bits of space and supports in 
constant time queries $access(S,i)$, $rank(S,i)$ if $S[i]=1$, and 
$select_1(S,j)$.

For general sequences, the wavelet tree \cite{GGV03} requires $N\lg n
+ o(N)$ bits of space \cite{GRR08} and supports all three operations
in $O(\lg n)$ time. Another representation, by Barbay et
al.~\cite{BGNN10}, requires at most $N\lg n + o(N\lg n)$ bits and
solves $access(S,i)$ in constant time and $select(S,j)$ in time
$O(\lg\lg n)$, or vice versa. Query $rank(S,i)$ takes time $O(\lg\lg
n)$.

\subsection{Labeled Binary Relations}
\label{sec:lbinrel}

A labeled binary relation corresponds to a binary relation $\Rel
\subseteq A\times B$, where $A=[1..n_1]$ and $B=[1..n_2]$, augmented
with a function $\Lab:A\times B \rightarrow L\cup\{\perp\}$,
$L=[1..\ell]$, that defines labels for each pair in $\Rel$, and
$\perp$ for pairs that are not in $\Rel$. Let us identify $A$ with the
columns and $B$ with the rows in a table. We describe a simplification
of a representation of binary relations \cite{CN09,CN11}, for the case
of this paper where each element of $A$ is associated to exactly one
element of $B$, so $|\Rel|=n_1$. We use a string $S_B[1..n_1]$ over
alphabet $[1..n_2]$, where $S_B[i]$ is the element of $B$ associated
to column $i$. A second string $S_\Lab[1..n_1]$ on alphabet
$[1..\ell]$ is stored, so that $S_\Lab[i]$ is the label corresponding
to the pair represented by $S_B[i]$.

If we use a wavelet tree for $S_B$ (see Section~\ref{sec:seqs}) and a plain 
string representation for $S_\Lab$, the total space is 
$n_1(\lg n_2 + \lg \ell) + O(n_1)$ bits. With this representation we can 
answer, among others, the following queries of interest in this paper.

\begin{itemize}
 \item Find the label of the element $b$ associated to a given $a$,
$S_\Lab[a]$, in $O(1)$ time.
 \item Enumerate the $k$ pairs
$(a,b)\in\Rel$ such that $a_1\leq a \leq a_2$ and $b_1\leq b \leq b_2$, in
$O((k+1)\lg n_2)$ time.
\end{itemize}

\subsection{Succinct Tree Representations}
\label{sec:trees}

There are many succinct tree representations for trees $\Tree$ with $N$ nodes. 
Most take $2N+o(N)$ bits of space. In this paper we use one called DFUDS
\cite{BDM+05}, which in particular answers in constant time the following
operations. Node identifiers $v$ are associated to a position in $[1..2N]$.

\begin{itemize}
\item $node(p)$: the node with preorder number $p$.   
\item $preorder(v)$: the preorder number of node $v$. 
\item $\mathit{leafrank}(v)$: number of leaves to the left of $v$. 
\item $\mathit{numleaves}(v)$: number of leaves below $v$. 
\item $parent(v)$: the parent of $v$. 
\item $child(v,k)$: the $k$th child of $v$.
\item $\mathit{nextsibling}(v)$: the next sibling of $v$. 
\item $degree(v)$: the number of children of $v$. 
\item $depth(v)$: the depth of $v$. 
\item $\levelanc(v,k)$: the $k$th ancestor of $v$. 
\end{itemize}

The DFUDS representation is obtained by traversing the tree in DFS
order and appending to a bitmap the degree of each node, written in
unary.  

\section{Preprocessing and Representing the Grammar}
\label{sec:preproc}

Let $\G$ be a grammar that generates a single string $T[1..u]$, formed
by $n$ (terminal and nonterminal) symbols. The $\sigma \le n$ terminal
symbols come from an alphabet $\Sigma=[1,\sigma],$\footnote{
Non-contiguous alphabets can be handled with some extra space, as shown
in previous work~\cite{CN11}.} and then $\G$ contains $n-\sigma$ rules
of the form $X_i \rightarrow \alpha_i$, one per nonterminal.  This
$\alpha_i$ is called the {\em right-hand side} of the rule. We call $N
= \sum |\alpha_i|$ the {\em size} of $\G$.  Note it holds $\sigma \le
N$, as the terminals must appear in the right-hand sides.  We assume
all the nonterminals are used to generate the string; otherwise unused
rules can be found and dropped in $O(N)$ time.

We preprocess $\G$ as follows.  First, for each terminal symbol
$a\in\Sigma$ present in $\G$ we create a rule $X_a \rightarrow a$, and
replace all other occurrences of $a$ in the grammar by $X_a$. As a
result, the grammar contains exactly $n$ nonterminal symbols $\X =
\{X_1, \ldots, X_n\}$, each associated to a rule $X_i \rightarrow
\alpha_i$, where $\alpha_i \in \Sigma$ or $\alpha_i \in \X^+$. We
assume that $X_n$ is the start symbol.  

Any rule $X_i \rightarrow \alpha_i$ where $|\alpha_i| \le 1$ (except
for $X_a \rightarrow a$) is removed by replacing $X_i$ by $\alpha_i$
everywhere, decreasing $n$ and without increasing $N$.

We further preprocess $\G$ to enforce the property that any
nonterminal $X_i$, except $X_n$ and those $X_i \rightarrow a \in
\Sigma$, must be mentioned in at least two right-hand sides. We
traverse the rules of the grammar, count the occurrences of each
symbol, and then rewrite the rules, so that only the rules of those
$X_i$ appearing more than once (or the excepted symbols) are
rewritten, and as we rewrite a right-hand side, we replace any
(non-excepted) $X_i$ that appears once by its right-hand side
$\alpha_i$. This transformation takes $O(N)$ time and does not alter
$N$ (yet it may reduce $n$).

Note that $n$ is now the number of rules in the transformed grammar $\G$. 
We will still call $N$ the size of the original grammar (the transformed
one has size at most $N+\sigma$).

We call $\F(X_i)$ the single string generated by $X_i$, that is
$\F(X_i)=a$ if $X_i \rightarrow a$ and $\F(X_i) = \F(X_{i_1})\ldots
\F(X_{i_k})$ if $X_i \rightarrow X_{i_1} \ldots X_{i_k}$.  $\G$
generates the text $T = \mathcal{L}(\G)=\F(X_n)$.

Our last preprocessing step, and the most expensive one, is to
renumber the nonterminals so that $i<j \Leftrightarrow
\F(X_i)^{rev}<\F(X_j)^{rev}$, where $S^{rev}$ is string $S$ read
backwards.  The usefulness of this reverse lexicographic order will be
apparent later.  The sorting can be done in time $O(u+n\lg n)$ and
$O(u\lg u)$ bits of space \cite{CN09,CN11}, which dominates the
previous time complexities. Let us say that $X_n$ became $X_s$ after the
reordering.

\medskip

We define now a structure that will be key in our index. 

\begin{definition}
The {\em grammar tree} of $\G$ is a general tree $\Tree_\G$ with nodes
labeled in $\X$. Its root is labeled $X_s$. Let $\alpha_s =
X_{s_1}\ldots X_{s_k}$. Then the root has $k$ children labeled
$X_{s_1},\ldots,X_{s_k}$. The subtrees of these children are defined
recursively, left to right, so that the first time we find a symbol
$X_i$, we define its children using $\alpha_i$. However, the next
times we find a symbol $X_i$, we leave it as a leaf of the grammar
tree (if we expanded it the resulting tree would be the {\em parse tree} of
$T$, with $u$ nodes). Also symbols $X_a \rightarrow a$ are not
expanded but left as leaves. We say that $X_i$ is {\em defined} in the
only internal node of $\Tree_\G$ labeled $X_i$.
\end{definition}

Since each right-hand side $\alpha_i \not= a \in\Sigma$ is written
once in the tree, plus the root $X_s$, the total number of nodes in
$\Tree_\G$ is $N+1$.

The grammar tree partitions $T$ in a way that is useful for finding
occurrences, using a concept that dates back to K\"arkk\"ainen
\cite{Kar99}, who used it for Lempel-Ziv parsings.

\begin{definition} Let $X_{l_1}, X_{l_2},\ldots$ be the nonterminals
labeling the consecutive leaves of $\Tree_\G$. Let $T_i =
\F(X_{l_i})$, then $T = T_1 T_2 \ldots$ is a partition of $T$
according to the leaves of $\Tree_\G$. An occurrence of pattern $P$ in
$T$ is called {\em primary} if it spans more than one $T_i$, and {\em
secondary} if it is inside some $T_i$.
\end{definition}

\begin{figure}
\begin{center}
\includegraphics[scale=0.58]{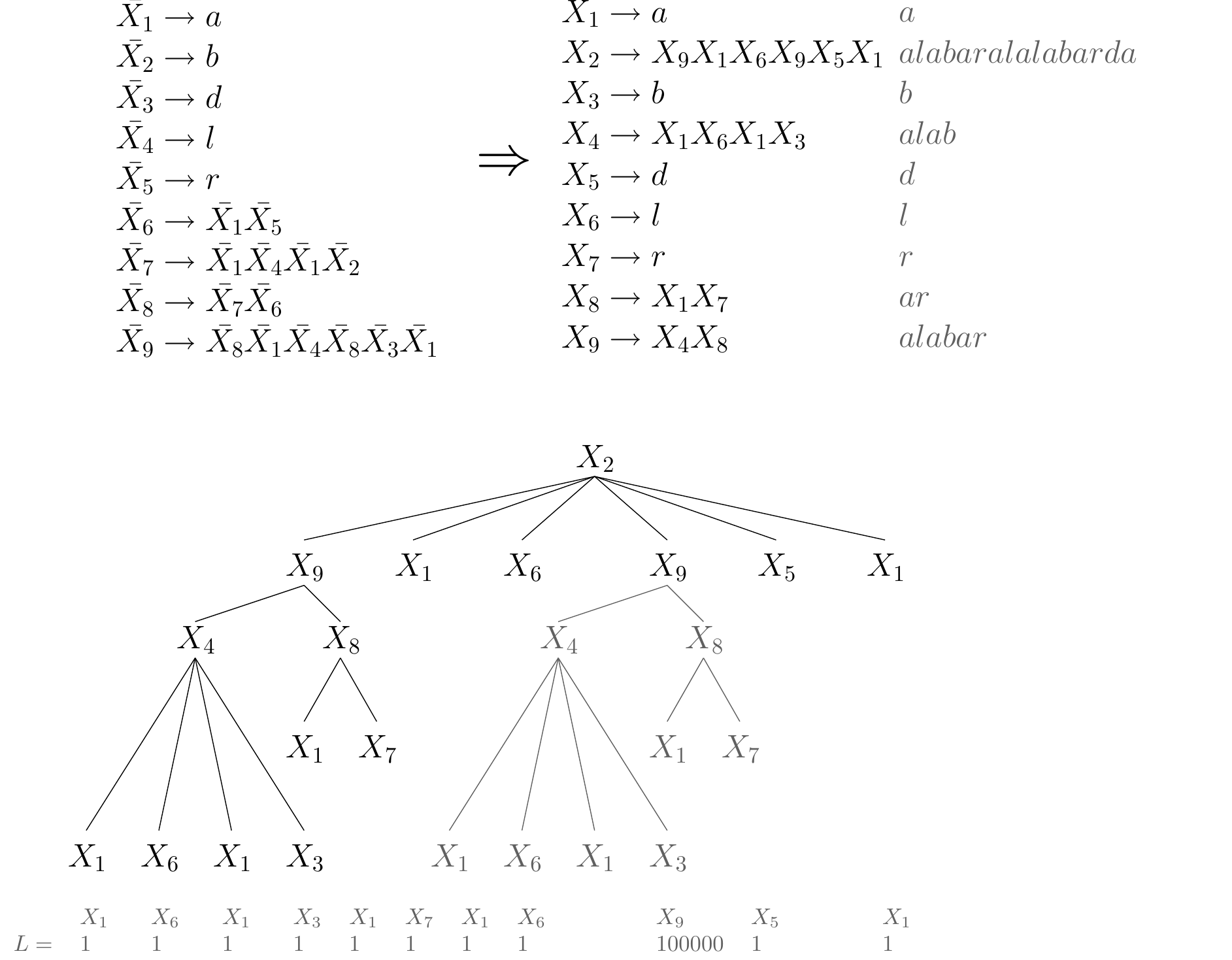}
\caption{On top left, a grammar $\G$ generating string
``{\em alabaralalabarda}''. On top right, our reordering of the grammar and
strings $\F(X_i)$. In the middle, the grammar tree $\Tree_\G$ in black; the 
whole parse tree includes also the grayed part. On the bottom we show our 
bitmap $L$ (Section~\ref{sec:extractarbitrary}).}
\label{fig:example_grammar}
\end{center}
\end{figure}

Figure \ref{fig:example_grammar} shows the reordering and grammar tree for a
grammar generating the string ``{\em alabaralalabarda}''.

Our self-index will represent $\G$ using two main components. A first
one represents the grammar tree $\Tree_\G$ using a DFUDS
representation (Section~\ref{sec:trees}) and a sequence of labels
(Section~\ref{sec:seqs}).  This will be used to extract text and
decompress rules. When augmented with a secondary trie $\Tree_S$
storing leftmost/rightmost paths in $\Tree_\G$, the representation will expand
any prefix/suffix of a rule in optimal time \cite{GKPS05}.

The second component in our self-index corresponds to a labeled binary
relation (Section~\ref{sec:lbinrel}), where $B=\X$ and $A$ is the set
of proper suffixes starting at positions $j+1$ of rules $\alpha_i$: 
$(\alpha_i[j],\langle
i,j+1\rangle)$ will be related for all $X_i \rightarrow \alpha_i$ and
$1 \le j < |\alpha_i|$.  This binary relation will be used to find the
primary occurrences of the search pattern. Secondary occurrences will
be tracked in the grammar tree.

\section{Extracting Text}
\label{sec:extract}

We first describe a simple structure that extracts the prefix of
length $\ell$ of any rule in $O(\ell+h)$ time.  We then augment this
structure to support extracting any substring of length $\ell$ in time
$O(\ell+h\lg(N/h))$, and finally augment it further to retrieve the
prefix or suffix of any rule in optimal $O(\ell)$ time. This last
result is fundamental for supporting searches, and is obtained by
extending the structure proposed by Gasieniec et al.~\cite{GKPS05} for
SLPs to general context-free grammars generating one
string.  The improvement does not work for extracting arbitrary
substrings, as in that case one has to find first the nonterminals
that must be expanded. This subproblem is not easy, especially in
little space \cite{BLRSSW11}.

As anticipated, we represent the topology of the grammar tree
$\Tree_\G$ using DFUDS \cite{BDM+05}. The sequence of labels
associated to the tree nodes is stored in preorder in a sequence
$X[1..N+1]$, using the fast representation of Section~\ref{sec:seqs}
where we choose constant time for $access(X,i)=X[i]$ and $O(\lg\lg n)$
time for $select_a(X,j)$.

We also store a bitmap $Y[1..n]$ that marks the rules of the form
$X_i\rightarrow a \in \Sigma$ with a 1-bit. Since the rules have been
renumbered in (reverse) lexicographic order, every time we find a rule
$X_i$ such that $Y[i]=1$, we can determine the terminal symbol it
represents as $a=rank_1(Y,i)$ in constant time.  In our example of
Figure~\ref{fig:example_grammar} this vector is $Y=101011100$.

\subsection{Expanding Prefixes of Rules}
\label{sec:expandprefix}

Expanding a rule $X_i$ that does not correspond to a terminal
is done as follows. By the definition of
$\Tree_\G$, the first left-to-right occurrence of $X_i$ in sequence
$X$ corresponds to the definition of $X_i$; all the rest are leaves in
$\Tree_\G$. Therefore, $v=node(select_{X_i}(X,1))$ is the node in
$\Tree_\G$ where $X_i$ is defined. We traverse the subtree rooted at
$v$ in DFS order. Every time we reach a leaf $u$, we compute its label
$X_j=X[preorder(u)]$, and either output a terminal if $Y[j]=1$ or
recursively expand $X_j$. This is in fact a traversal of the {\em
parse tree} starting at node $v$, using instead the grammar tree. Such
a traversal takes $O(\ell+h_v)$ steps \cite{CN09,CN11}, where $h_v \le
h$ is the height of the parsing subtree rooted at $v$. In particular,
if we extract the whole rule $X_i$ we pay $O(\ell)$ steps, since we
have removed unary paths in the preprocessing of $\G$ and thus $v$ has
$\ell > h_v$ leaves in the parse tree. The only obstacle to having 
constant-time steps are the queries $select_{X_i}(X,1)$. As these are only 
for the position 1, we can have them precomputed in a sequence $F[1..n]$
using $n \lceil \lg N\rceil = n\lg n + O(N)$ further bits of space.

The total space required for $\Tree_\G$, considering the DFUDS
topology, sequence $X$, bitmap $Y$, and sequence $F$, is
$N\lg n + n\lg n + o(N\lg n)$ bits. We reduce the space to 
$N\lg n + \delta\, n\lg n + o(N\lg n)$, for any $0<\delta\le 1$,
as follows. Form a sequence $X'[1..N-n+1]$ where the first position 
of every symbol $X_i$ in $X$ has been removed, and mark in a bitmap
$Z[1..N+1]$, with a $1$, those first positions in $X$. Replace our sequence $F$
by a permutation $\pi[1..n]$
so that $select_{X_i}(X,1) = F[i] = select_1(Z,\pi[i])$. Now we can still 
access any $X[i] = X'[rank_0(Z,i)]$ if $Z[i]=0$. For the case $Z[i]=1$ we 
have $X[i] = \pi^{-1}[rank_1(Z,i)]$. Similarly, 
$select_{X_i}(X,j) = select_0(Z,select_{X_i}(X',j-1))$ for $j>1$. 
Then use $Z$, $\pi$, and $X'$ instead of $F$ and $S$.

All the operations retain the same times except for the access to $\pi^{-1}$. 
We use for $\pi$ a representation by Munro et al.~\cite{MRRR03} that takes 
$(1+\delta)n\lg n$ bits and computes any $\pi[i]$ in constant time and 
any $\pi^{-1}[j]$ in time $O(1/\delta)$, which will be the cost to access
$X$. Although this will have an impact later, we note that for extraction we
only access $X$ at leaf nodes, where it always takes constant time.%
\footnote{Nonterminals $X_a \rightarrow a$ do not have a definition in 
$\Tree_\G$, so they are not extracted from $X$ nor
represented in $\pi$, thus they are accessed in constant time. They can be 
skipped from $\pi[1..n]$ with bitmap $Y$, so that in fact $\pi$ is of length
$n-\sigma$ an is accessed as $\pi[rank_0(Y,i)]$; for $\pi^{-1}$ we actually
use $select_0(Y,\pi^{-1}[j]$).}

\subsection{Extracting Arbitrary Substrings}
\label{sec:extractarbitrary}

In order to extract any given substring of $T$, we add a bitmap
$L[1..u+1]$ that marks with a 1 the first position of each $T_i$ in
$T$ (see Figure~\ref{fig:example_grammar}).  We can then compute the
starting position of any node $v \in \Tree_\G$ as
$select_1(L,\mathit{leafrank}(v)+1)$.

To extract $T[p,p+\ell-1]$, we binary search the starting position $p$
from the root of $\Tree_\G$. If we arrive at a leaf that does not
represent a terminal, we go to its definition in $\Tree_\G$, translate
position $p$ to the area below the new node $v$, and continue
recursively. At some point we finally reach the position $p$, and from
there on we extract the symbols rightwards. Just as before, the total
number of steps is $O(\ell+h)$. However, the $h$ steps require binary
searches. As there are at most $h$ binary searches among the children
of different tree nodes, and there are $N+1$ nodes, in the worst case
the binary searches cost $O(h \lg (N/h))$, thus the total cost is
$O(\ell + h \lg (N/h))$.

The number of ones in $L$ is at most $N$. Since we only need
$select_1$ on $L$, we can use an ID representation (see
Section~\ref{sec:seqs}), requiring $N\lg(u/N)+O(N+\lg\lg u) =
N\lg(u/N)+O(N)$ bits (since $N \ge \lg u$ in any grammar). Thus the
total space becomes $N\lg n + N\lg(u/N) + \delta\, n\lg n + o(N\lg n)$ bits.

\subsection{Optimal Expansion of Rule Prefixes and Suffixes}

Our improved version builds on the proposal by Gasieniec et
al.~\cite{GKPS05}. We show how to extend their representation using
succinct data structures so that we can handle general grammars
instead of only SLPs. Following their notation, call
$S(X_i)$ the string of labels of the nodes in the path from any node
labeled $X_i$ to its leftmost leaf in {\em the parse tree} (we take as
leaves the nonterminals $X_a \in \X$, not the terminals $a\in\Sigma$).
We insert all the strings $S(X_i)^{rev}$ into a trie $\Tree_S$. Note
that each symbol appears only once in $\Tree_S$ \cite{GKPS05}, thus it
has $n$ nodes. Again, we represent the topology of $\Tree_S$ using
DFUDS. However, its sequence of labels $X_S[1..n]$ turns out to be a
permutation in $[1..n]$, for which we use again the representation by 
Munro et al.~\cite{MRRR03} that takes $(1+\epsilon)n\lg n$ bits and 
computes any $X_S[i]$ in constant time and any $X_S^{-1}[j]$ in time
$O(1/\epsilon)$.

\begin{figure}
\begin{center}
\includegraphics[scale=0.6]{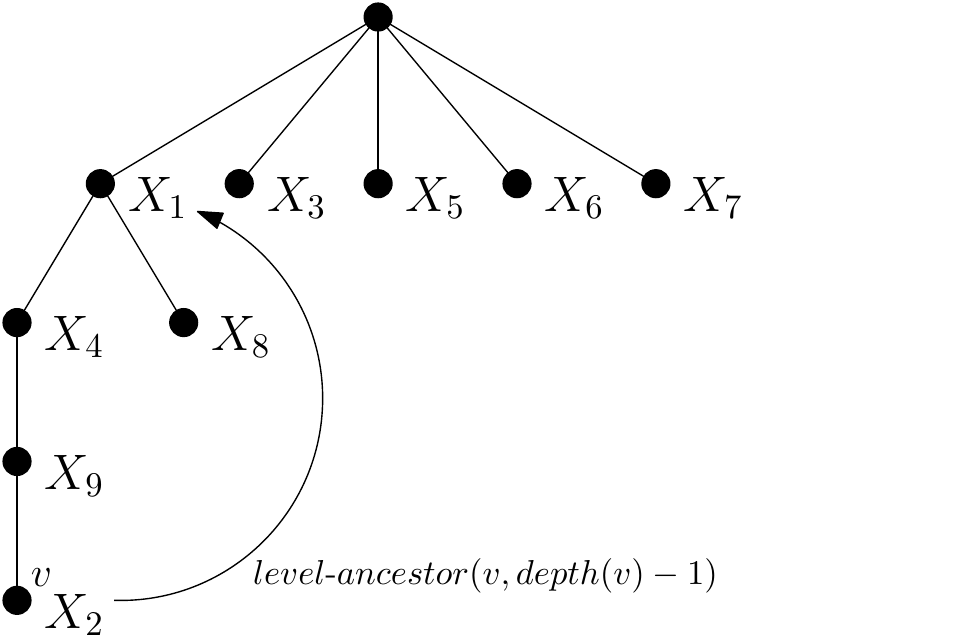}
\caption{Example of the trie of leftmost paths for the grammar of
Figure~\ref{fig:example_grammar}. The arrow pointing from $X_2$ to
$X_1$ ilustrates the procedure to determine the first terminal symbol
generated by $X_2$.}
\label{fig:example_trie}
\end{center}
\end{figure}

We can determine the first terminal in the expansion of $X_i$, which
labels node $v \in \Tree_S$, as follows. Since the last symbol in
$S(X_i)$ is a nonterminal $X_a$ representing some $a \in \Sigma$, it
follows that $X_i$ descends in $\Tree_S$ from $X_a$, which is a child
of the root.  This node is $v_a = \levelanc(v,depth(v)-1)$. Then $a =
rank_1(Y,X_S[preorder(v_a)])$. Figure \ref{fig:example_trie} shows an
example of this particular query in the trie for the grammar presented
in Figure \ref{fig:example_grammar}.

A prefix of $X_i$ is extracted as follows. First, we obtain the
corresponding node $v \in \Tree_S$ as $v = X_S^{-1}[X_i]$. Then we
obtain the leftmost symbol of $v$ as explained. The remaining symbols
descend from the second and following children, in the parse tree, of
the nodes in the upward path from a node labeled $X_i$ to its leftmost
leaf, or which is the same, of the nodes in the downward path from the
root of $\Tree_S$ to $v$. Therefore, for each node $w$ in the list 
$\levelanc(v,depth(v)-2), \ldots,parent(v),v$, we map $w$ to its definition 
$u \in \Tree_\G$, $u = node(select_{X_j}(X,1))$ where $X_j = X_S[preorder(w)]$.
Once $u$ is found, we recursively expand its children, from the second to the
last, by mapping them back to $\Tree_S$, and so on. By charging the
cost to the new symbol to expand, and because there are no unary
paths, it can be seen that we carry out $O(\ell)$ steps to extract the
first $\ell$ symbols.  Moreover, the extraction is real-time
\cite{GKPS05}.  All costs per step are constant except for the $O(1/\epsilon)$
to access $X_S^{-1}$.

For extracting suffixes of rules in $\G$, we need another version of
$\Tree_S$ that stores the rightmost paths. This leads to our first
result (choosing $\delta=o(1)$).

\begin{lemma} \label{lem:extract} Let a sequence $T[1..u]$ be
represented by a context free grammar with $n$ symbols, size $N$, and
height $h$. Then, for any $0<\epsilon \le 1$, there exists a data
structure using at most $N\lg n + N\lg(u/N) + (2+\epsilon)n\lg n +
o(N\lg n)$ bits of space that extracts any substring of length $\ell$
from $T$ in time $O(\ell+h\lg(N/h))$, and a prefix or suffix of length
$\ell$ of the expansion of any nonterminal in time $O(\ell/\epsilon)$.
\end{lemma}

\section{Locating Patterns}

A secondary occurrence of $P$ inside a leaf of $\Tree_\G$ labeled by a
symbol $X_i$ occurs as well in the internal node of $\Tree_G$ where
$X_i$ is defined. If that occurrence is also secondary, then it occurs
inside a child $X_j$ of $X_i$, and we can repeat the argument with
$X_j$ until finding a primary occurrence inside some $X_k$. This shows
that all the secondary occurrences can be found by first spotting the
primary occurrences, and then finding all the copies of the
nonterminal $X_k$ that contain the primary occurrences, as well as all
the copies of the nonterminals that contain $X_k$, recursively.

The strategy \cite{Kar99} to find the primary occurrences of
$P=p_1p_2\ldots p_m$ is to consider the $m-1$ partitions $P = P_1
\cdot P_2$, $P_1 = p_1 \ldots p_i$ and $P_2 = p_{i+1}\ldots p_m$, for
$1 \le i < m$. For each partition we will find all the nonterminals
$X_k \rightarrow X_{k_1} X_{k_2} \ldots X_{k_r}$ such that $P_1$ is a
suffix of some $\F(X_{k_i})$ and $P_2$ is a prefix of
$\F(X_{k_{i+1}})\ldots \F(X_{k_r})$. This finds each primary
occurrence exactly once. The secondary occurrences are then tracked in
the grammar tree $\Tree_\G$.\footnote{If $m=1$ we can just find all
the occurrences of $X_{p_1}$ in $\Tree_\G$ and track its secondary
occurrences.}

\subsection{Finding Primary Occurrences}

As anticipated at the end of Section~\ref{sec:preproc}, we store a
binary relation $\Rel \subseteq A \times B$ to find the primary
occurrences. It has $n$ rows labeled $X_i$, for all $X_i \in \X = B$,
and $N-n$ columns. Each column, denoted $\langle i,j+1\rangle$,
corresponds to a distinct proper suffix $\alpha_i[j+1..]$ of a
right-hand side $\alpha_i$.  The labels belong to $[1..N+1]$. The
relation contains one pair per column: $(\alpha_i[j],\langle
i,j+1\rangle) \in \Rel$ for all $1 \le i \le n$ and $1 \le j <
|\alpha_i|$. Its label is the preorder of the $(j+1)$th child of the
node $v \in \Tree_\G$ where $X_i$ is defined. The space for the binary
relation is $(N-n)(\lg n+\lg N)+O(N)$ bits.

\begin{figure}
\begin{center}
\includegraphics[scale=0.6]{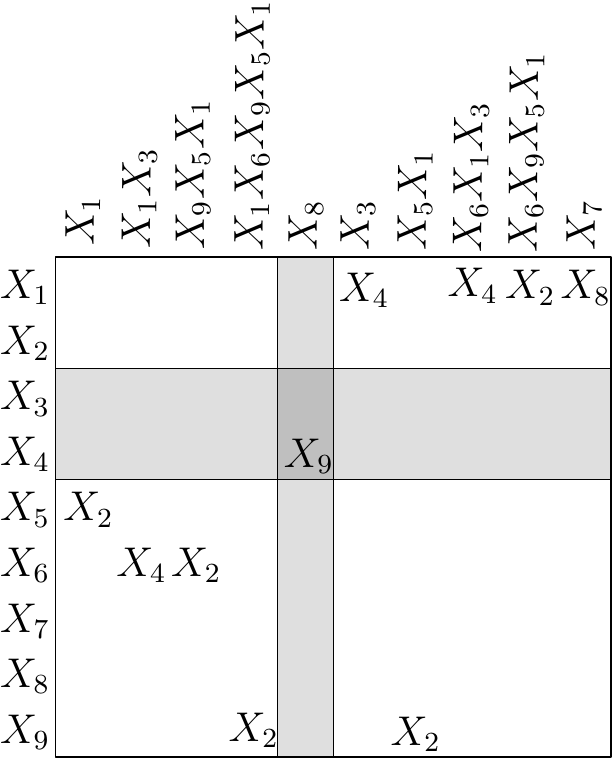}
\caption{Relation $\Rel$ for the grammar presented in Figure
\ref{fig:example_grammar}. The highlighted ranges correspond to the
result of searching for $b\cdot ar$, where the single primary occurrence
corresponds to $X_2$.}
\label{fig:binrel}
\end{center}
\end{figure}

Recall that, in our preprocessing, we have sorted $\X$ according to
the lexicographic order of $\F(X_i)^{rev}$. We also sort all the pairs
$\langle i,j+1 \rangle$ lexicographically according to the suffixes
$\F(\alpha_i[j+1])\F(\alpha_i[j+2]) \ldots \F(\alpha_i[|\alpha_i|])$.
This can be done in $O(u+N\lg N)$ time in a way similar to how $\X$
was sorted: Each pair $\langle i,j+1\rangle$, labeled $p$, can be
associated to the substring
$T[select_1(L,\mathit{rankleaf}(node(p))+1) \ldots
select_1(L,\mathit{rankleaf}(v)+\mathit{numleaves}(v)+1)-1]$, where
$v$ is the parent of $node(p)$. Then we can proceed as in previous
work \cite{CN09,CN11}. Figure \ref{fig:binrel} illustrates how
$\Rel$ is used, for the grammar presented in Figure
\ref{fig:example_grammar}.

Given $P_1$ and $P_2$, we first find the range of rows
whose expansions finish with $P_1$, by binary searching for $P_1^{rev}$ in
the expansions $\F(X_i)^{rev}$. Each comparison in the binary search
needs to extract $|P_1|$ terminals from the suffix of
$\F(X_i)$. According to Lemma~\ref{lem:extract}, this takes
$O(|P_1|/\epsilon)$ time. Similarly, we binary search for the range of
columns whose expansions start with $P_2$. Each comparison needs to
extract $\ell=|P_2|$ terminals from the prefix of
$\F(\alpha_i[j+1])\F(\alpha_i[j+2])\ldots$. Let $r$ be the column we
wish to compare to $P_2$. We extract the label $p$ associated to the
column in constant time (recall Section~\ref{sec:lbinrel}). Then we
extract the first $\ell$ symbols from the expansion of $node(p) \in
\Tree_\G$. If $node(p)$ does not have enough symbols, we continue with
$\mathit{nextsibling}(p)$, and so on, until we extract $\ell$ symbols
or we exhaust the suffix of the rule.  According to
Lemma~\ref{lem:extract}, this requires time $O(|P_2|/\epsilon)$.  Thus
our two binary searches require time $O((m/\epsilon)\lg N)$.

This time can be further improved by using the same technique as in
previous work \cite{CN11}. The idea is to sample phrases at regular
intervals and store the sampled phrases in a Patricia tree
\cite{Mor68}.  We first search for the pattern in the Patricia tree,
and then complete the process with a binary search between two sampled
phrases (we first verify the correctness of the Patricia search by
checking that our pattern is actually within the range found).  By
sampling every $ \lg u \lg\lg n / \lg n$ phrases, the resulting time
for searching becomes $O\left(m\lg\left(\frac{\lg u}{\lg
n}\right)\right)$ and we only require $o(N\lg n)$ bits of extra space,
as the Patricia tree needs $O(\lg u)$ bits per node.

Once we have identified a range of rows $[a_1,a_2]$ and a range of
columns $[b_1,b_2]$, we retrieve all the points in the rectangle and
their labels, each in time $O(\lg n)$, according to
Section~\ref{sec:lbinrel}.  The parents of all the nodes $node(p) \in
\Tree_\G$, for each point $p$ in the range, correspond to the primary
occurrences. In Section 5.2 we show how to report primary and
secondary occurrences starting directly from those $node(p)$
positions.

Recall that we have to carry out this search for $m-1$ partitions of
$P$, whereas each primary occurrence is found exactly once. Calling
$occ$ the number of primary occurrences, the total cost of this part
of the search is $O\left((m^2/\epsilon)\lg \left(\frac{\lg u}{\lg
n}\right) + occ \lg n\right)$.

\subsection{Tracking Occurrences Through the Grammar Tree}

The remaining problem is how to track all the secondary occurrences
triggered by a primary occurrence, and how to report the positions
where these occur in $T$. Given a primary occurrence for partition
$P=P_1\cdot P_2$ located at $u=node(p) \in \Tree_\G$, we obtain the
starting position of $P$ in $T$ by moving towards the root while
keeping count of the offset between the beginning of the current node
and the occurrence of $P$. Initially, for node $u$ itself, this is $l
= -|P_1|$. Now, as long as $u$ is not the root, we set $l \leftarrow l
+ select_1(L,\mathit{rankleaves}(u)+1) -
select_1(L,\mathit{rankleaves}(parent(u))+1)$ and then $u \leftarrow
parent(u)$.  When we arrive at the root, the occurrence of $P$ starts
at $l$.

It seems like we are doing this $h$ times in the worst case, since we
need to track the occurrence up to the root. In fact we might do so
for some symbols, but the total cost is amortized. Evey time we move
from $u$ to $v=parent(u)$, we know that $X[v]$ appears at least once
more in the tree. This is because of our preprocessing
(Section~\ref{sec:preproc}), where we force rules to appear at least
twice or be removed. Thus $v$ defines $X[v]$, but there are one or
more leaves labeled $X[v]$, and we have to report the occurrences of
$P$ inside them all. For this sake we carry out $select_{X[v]}(X,i)$
for $i=1,2\ldots$ until spotting all those occurrences (where $P$
occurs with the current offset $l$). We recursively track them to the
root of $\Tree_\G$ to find their absolute position in $T$, and
recursively find the other occurrences of all their ancestor
nodes. The overall cost amortizes to $O(1)$ steps per occurrence
reported, as we can charge the cost of moving from $u$ to $v$ to the
other occurrence of $v$. If we report $occ$ secondary occurrences we
carry out $O(occ)$ steps, each costing $O(\lg\lg n)$ time. We can thus
use $\delta=O(1/\lg\lg n)$ (Section~\ref{sec:expandprefix}) so that 
the cost to access $X[v]$ does not
impact the space nor time complexity.

By adding up the space of Lemma~\ref{lem:extract} with that of the
labeled binary relation, and adding up the costs, we have our central
result, Theorem~\ref{thm:main}.

\section{Conclusions}

We presented the first grammar-based text index whose locate time does
not depend on the height of the grammar. There are previous results on
generating balanced grammars to compress text, as for example the ones
proposed by Rytter \cite{Ryt03} and Sakamoto \cite{Sak05}. These
representations allow previous indexing techniques to guarantee
sublinear locating times, yet these techniques introduce a penalty in
the size of the grammar.  Our index also extends the grammar-based
indexing techniques to a more general class of grammars than SLPs, the
only class explored so far in this scenario.

We note that in our index each primary occurrence is reported in
$O(\lg n)$ time, whereas each secondary ones requires just $O(\lg\lg
n)$ time. The complexity of primary occurrences is dominated by the
time to report points in a range using our binary relation
representation. We believe this can be lowered up to $O(\lg\lg n)$, in
exchange for using more space. For example, Bose et al.~\cite{BHMM09}
represent an $n \times n$ grid with $n$ points within $n\lg n + o(n\lg
n)$ bits, so that each point in a range can be reported in time $O(\lg
n / \lg\lg n)$; using $O((1/\epsilon)n\lg n)$ bits the time can be
reduced to $O(\lg^\epsilon n)$ for any constant $0<\epsilon\le 1$
\cite{Cha88,Nek09,CLP11}; and using $O(n \lg n \lg\lg n)$ bits one can
achieve time $O(\lg\lg n)$ \cite{ABR00,CLP11} (all these solutions
have a small additive time that is not relevant for our
application). It seems likely that these structures can be extended to
represent our $n \times N$ grid with $N$ points (i.e., our string
$S_B$).  In the case of Bose et al.\ this could even be asymptotically
free in terms of space.

Alternatively, instead of speeding up the reporting of primary
occurrences, we can slow down that of secondary occurrences so that
they match, and in exchange reduce the space. For example, one of our
largest terms in the index space owes to the need of storing the
phrase lengths in $\Tree_\G$. By storing just the $n$ internal node
lengths and one out of $\lg n$ lengths at the leaves of $\Tree_\G$, we
reduce $N\lg(u/N)$ bits of space in our index to $(n+(N-n)/\lg
n)\lg(u/(n+(N-n)/\lg n)) \le (n+N/\lg n)\lg(u/N) + o(N\lg n)$. Note
this penalizes the extraction time by an $O(\lg n)$ factor in the
worst case.

Several questions remain open, for example: Is it possible to lower
the dependence on $m$ to linear, as achieved in some LZ78-based
schemes \cite{RO08}? Is it possible to reduce the space to $N\lg n +
o(N\lg n)$, that is, asymptotically the same as the compressed text,
as achieved on statistical-compression-based self-indexes \cite{NM07}?
Is it possible to remove $h$ from the extraction complexity within
less space than the current solutions \cite{BLRSSW11}?

\bibliography{paper}

\begin{thebibliography}{10}

\bibitem{ABR00}
S.~Alstrup, G.~Brodal, and T.~Rauhe.
\newblock New data structures for orthogonal range searching.
\newblock In {\em Proc. 41st Annual Symposium on Foundations of Computer
  Science (FOCS)}, pages 198--207, 2000.

\bibitem{AB92}
A.~Amir and G.~Benson.
\newblock Efficient two-dimensional compressed matching.
\newblock In {\em Proc. 2nd Data Compression Conference (DCC)}, pages 279--288,
  1992.

\bibitem{ANScpm06.2}
D.~Arroyuelo, G.~Navarro, and K.~Sadakane.
\newblock Reducing the space requirement of {LZ}-index.
\newblock In {\em Proc. 17th Annual Symposium on Combinatorial Pattern Matching
  (CPM)}, LNCS 4009, pages 319--330, 2006.

\bibitem{BGNN10}
J.~Barbay, T.~Gagie, G.~Navarro, and Y.~Nekrich.
\newblock Alphabet partitioning for compressed rank/select and applications.
\newblock In {\em Proc. 21st Annual International Symposium on Algorithms and
  Computation (ISAAC)}, LNCS 6507, pages 315--326, 2010.
\newblock Part II.

\bibitem{BDM+05}
D.~Benoit, E.~Demaine, I.~Munro, R.~Raman, V.~Raman, and S.~Srinivasa Rao.
\newblock Representing trees of higher degree.
\newblock {\em Algorithmica}, 43(4):275--292, 2005.

\bibitem{BLRSSW11}
Philip Bille, Gad~M. Landau, Rajeev Raman, Kunihiko Sadakane, Srinivasa~Rao
  Satti, and Oren Weimann.
\newblock Random access to grammar-compressed strings.
\newblock In {\em Proc. 22nd Annual ACM-SIAM Symposium on Discrete Algorithms
  (SODA'11)}, pages 373--389, 2011.

\bibitem{BHMM09}
P.~Bose, M.~He, A.~Maheshwari, and P.~Morin.
\newblock Succinct orthogonal range search structures on a grid with
  applications to text indexing.
\newblock In {\em Proc. 11th Algorithms and Data Structures Symposium (WADS)},
  LNCS 5664, pages 98--109, 2009.

\bibitem{CLP11}
T.~Chan, K.~Larsen, and M.~Patrascu.
\newblock Orthogonal range searching on the {RAM}, revisited.
\newblock {\em CoRR}, abs/1103.5510, 2011.

\bibitem{CLLPPSS05}
M.~Charikar, E.~Lehman, D.~Liu, R.~Panigrahy, M.~Prabhakaran, A.~Sahai, and
  A.~Shelat.
\newblock The smallest grammar problem.
\newblock {\em IEEE Transactions on Information Theory}, 51(7):2554--2576,
  2005.

\bibitem{Cha88}
B.~Chazelle.
\newblock Functional approach to data structures and its use in
  multidimensional searching.
\newblock {\em SIAM Journal on Computing}, 17(3):427--462, 1988.

\bibitem{CFMPN10}
F.~Claude, A.~Fari{\~n}a, M.~Mart{\'{\i}}nez-Prieto, and G.~Navarro.
\newblock Compressed $q$-gram indexing for highly repetitive biological
  sequences.
\newblock In {\em Proc. 10th IEEE Conference on Bioinformatics and
  Bioengineering (BIBE)}, 2010.

\bibitem{CFMPN11}
F.~Claude, A.~Fari{\~n}a, M.~Mart{\'{\i}}nez-Prieto, and G.~Navarro.
\newblock Indexes for highly repetitive document collections.
\newblock In {\em Proc. 20th ACM International Conference on Information and
  Knowledge Management (CIKM)}, 2011.
\newblock To appear.

\bibitem{CN09}
F.~Claude and G.~Navarro.
\newblock Self-indexed text compression using straight-line programs.
\newblock In {\em Proc. 34th International Symposium on Mathematical
  Foundations of Computer Science (MFCS)}, LNCS 5734, pages 235--246. Springer,
  2009.

\bibitem{CN11}
F.~Claude and G.~Navarro.
\newblock Self-indexed grammar-based compression.
\newblock {\em Fundamenta Informaticae}, 2011.
\newblock To appear. {\tt www.dcc.uchile.cl/gnavarro/ps/fi10.pdf}.

\bibitem{FM05}
P.~Ferragina and G.~Manzini.
\newblock Indexing compressed texts.
\newblock {\em Journal of the ACM}, 52(4):552--581, 2005.

\bibitem{GKPS05}
L.~Gasieniec, R.~Kolpakov, I.~Potapov, and P.~Sant.
\newblock Real-time traversal in grammar-based compressed files.
\newblock In {\em Proc. 15th Data Compression Conference (DCC)}, pages
  458--458, 2005.

\bibitem{GRR08}
A.~Golynski, R.~Raman, and S.~Rao.
\newblock On the redundancy of succinct data structures.
\newblock In {\em Proc. 11th Scandinavian Workshop on Algorithm Theory (SWAT)},
  LNCS 5124, pages 148--159, 2008.

\bibitem{GGV03}
R.~Grossi, A.~Gupta, and J.~Vitter.
\newblock High-order entropy-compressed text indexes.
\newblock In {\em Proc. 14th Annual ACM-SIAM Symposium on Discrete Algorithms
  (SODA)}, pages 841--850, 2003.

\bibitem{Kar99}
J.~K{\"a}rkk{\"a}inen.
\newblock {\em Repetition-Based Text Indexing}.
\newblock PhD thesis, Department of Computer Science, University of Helsinki,
  Finland, 1999.

\bibitem{KMSTSA03}
T.~Kida, T.~Matsumoto, Y.~Shibata, M.~Takeda, A.~Shinohara, and S.~Arikawa.
\newblock Collage system: a unifying framework for compressed pattern matching.
\newblock {\em Theoretical Computer Science}, 298(1):253--272, 2003.

\bibitem{KY00}
J.~Kieffer and E.-H. Yang.
\newblock Grammar-based codes: A new class of universal lossless source codes.
\newblock {\em IEEE Transactions on Information Theory}, 46(3):737--754, 2000.

\bibitem{KN11}
S.~Kreft and G.~Navarro.
\newblock Self-indexing based on {LZ77}.
\newblock In {\em Proc. 22th Annual Symposium on Combinatorial Pattern Matching
  (CPM)}, LNCS 6661, pages 41--54, 2011.

\bibitem{KBSCZ09}
S.~Kuruppu, B.~Beresford-Smith, T.~Conway, and J.~Zobel.
\newblock Repetition-based compression of large {DNA} datasets.
\newblock In {\em Proc. 13th Annual International Conference on Computational
  Molecular Biology (RECOMB)}, 2009.
\newblock Poster.

\bibitem{LM00}
J.~Larsson and A.~Moffat.
\newblock Off-line dictionary-based compression.
\newblock {\em Proceedings of the IEEE}, 88(11):1722--1732, 2000.

\bibitem{MNSV09}
V.~M{\"a}kinen, G.~Navarro, J.~Sir\'en, and N.~V{\"a}lim{\"a}ki.
\newblock Storage and retrieval of individual genomes.
\newblock In {\em Proc. 13th Annual International Conference on Computational
  Molecular Biology (RECOMB)}, LNCS 5541, pages 121--137, 2009.

\bibitem{Mor68}
D.~Morrison.
\newblock {PATRICIA} -- practical algorithm to retrieve information coded in
  alphanumeric.
\newblock {\em Journal of the ACM}, 15(4):514--534, 1968.

\bibitem{MRRR03}
J.~Munro, R.~Raman, V.~Raman, and S.~Srinivasa Rao.
\newblock Succinct representations of permutations.
\newblock In {\em Proc. 30th International Colloquium on Automata, Languages,
  and Programming (ICALP)}, LNCS 2719, pages 345--356, 2003.

\bibitem{NM07}
G.~Navarro and V.~M{\"a}kinen.
\newblock Compressed full-text indexes.
\newblock {\em ACM Computing Surveys}, 39(1):article 2, 2007.

\bibitem{Nek09}
Y.~Nekrich.
\newblock Orthogonal range searching in linear and almost-linear space.
\newblock {\em Computational Geometry: Theory and Applications},
  42(4):342--351, 2009.

\bibitem{NMWM04}
C.~Nevill-Manning, I.~Witten, and D.~Maulsby.
\newblock Compression by induction of hierarchical grammars.
\newblock In {\em Proc. 4th Data Compression Conference (DCC)}, pages 244--253,
  1994.

\bibitem{RRR02}
R.~Raman, V.~Raman, and S.~Rao.
\newblock Succinct indexable dictionaries with applications to encoding $k$-ary
  trees and multisets.
\newblock In {\em Proc. 13th Annual ACM-SIAM Symposium on Discrete Algorithms
  (SODA)}, pages 233--242, 2002.

\bibitem{RO08}
L.~Russo and A.~Oliveira.
\newblock A compressed self-index using a {Z}iv-{L}empel dictionary.
\newblock {\em Information Retrieval}, 11(4):359--388, 2008.

\bibitem{Ryt03}
W.~Rytter.
\newblock Application of {L}empel-{Z}iv factorization to the approximation of
  grammar-based compression.
\newblock {\em Theoretical Computer Science}, 302(1-3):211--222, 2003.

\bibitem{Sak05}
H.~Sakamoto.
\newblock A fully linear-time approximation algorithm for grammar-based
  compression.
\newblock {\em Journal of Discrete Algorithms}, 3:416--430, 2005.

\bibitem{ZL77}
J.~Ziv and A.~Lempel.
\newblock A universal algorithm for sequential data compression.
\newblock {\em IEEE Transactions on Information Theory}, 23(3):337--343, 1977.

\bibitem{ZL78}
J.~Ziv and A.~Lempel.
\newblock Compression of individual sequences via variable length coding.
\newblock {\em IEEE Transactions on Information Theory}, 24(5):530--536, 1978.

\end{thebibliography}

\end{document}